\documentclass[conference, a4paper]{IEEEtran}
\IEEEoverridecommandlockouts

\usepackage[noadjust]{cite}
\usepackage{amsmath,amssymb,amsfonts}
\usepackage{algorithmic}
\usepackage{graphicx}
\usepackage{textcomp}
\usepackage{xcolor}
\usepackage{siunitx}
\usepackage{stfloats}
\usepackage{booktabs}
\usepackage[percent]{overpic}
\usepackage{placeins}
\usepackage[capitalise]{cleveref}

\usepackage{fancyhdr}

\usepackage[utf8]{inputenc}

\newcolumntype{Y}{>{\centering\arraybackslash}X}
\usepackage[acronym]{glossaries}

\def\BibTeX{{\rm B\kern-.05em{\sc i\kern-.025em b}\kern-.08em
    T\kern-.1667em\lower.7ex\hbox{E}\kern-.125emX}}
    
\usepackage{eso-pic}
\usepackage{url}
\AddToShipoutPictureBG*{
  \AtPageUpperLeft{%
    \put(0,-40){\raisebox{15pt}{\makebox[\paperwidth]{\begin{minipage}{21cm}\centering
      \textcolor{gray}{This article has been accepted for publication in the proceedings of the 2022 29th IEEE International Conference on Electronics, Circuits \\and Systems (ICECS). Content may change prior to final publication. DOI: 10.1109/ICECS202256217.2022.9970865} 
    \end{minipage}}}}%
  }
  \AtPageLowerLeft{%
    \raisebox{25pt}{\makebox[\paperwidth]{\begin{minipage}{21cm}\centering
      \textcolor{gray}{ \copyright 2022 IEEE.  Personal use of this material is permitted.  Permission from IEEE must be obtained for all other uses, in any current or future media, including reprinting/republishing this material for advertising or promotional purposes, creating new collective works, for resale or redistribution to servers or lists, or reuse of any copyrighted component of this work in other works.
      }
    \end{minipage}}}%
  }
}

\begin{document}

\title{RF Power Transmission for Self-sustaining Miniaturized IoT Devices}


\definecolor{darkred}{HTML}{A8322C}
\newcommand{\revA}[1]{\textcolor{black}{#1}}

\author{
\IEEEauthorblockN{
Lukas Schulthess\IEEEauthorrefmark{1}, 
Federico Villani\IEEEauthorrefmark{1}, 
Philipp Mayer\IEEEauthorrefmark{1}, 
Michele Magno\IEEEauthorrefmark{1}}
\IEEEauthorblockA{\IEEEauthorrefmark{1}Dept. of Information Technology and Electrical Engineering, ETH Z\"{u}rich, Switzerland} 
}

\markboth{Journal of \LaTeX\ Class Files,~Vol.~14, No.~8, August~2015}%
{Shell \MakeLowercase{\textit{et al.}}: Bare Demo of IEEEtran.cls for IEEE Journals}

\maketitle

\begin{abstract}\label{sec:abstract}
Radio Frequency (RF) wireless power transfer is a promising technology that has the potential to constantly power small Internet of Things (IoT) devices, enabling even battery-less systems and reducing their maintenance requirements. However, to achieve this ambitious goal, carefully designed RF energy harvesting (EH) systems are needed to minimize the conversion losses and the conversion efficiency of the limited power. For intelligent internet of things sensors and devices, which often have non-constant power requirements, an additional power management stage with energy storage is needed to temporarily provide a higher power output than the power being harvested. 
This paper proposes an RF wireless power energy conversion system for miniaturized IoT composed of an impedance matching network, a rectifier, and power management with energy storage. The proposed sub-system has been experimentally validated and achieved an overall power conversion efficiency (PCE) of over \SI{30}{\percent} for an input power of \SI{-10}{dBm} and a peak efficiency of \SI{57}{\percent} at \SI{3}{dBm}.

{\color{blue}{
}}
\end{abstract}
\vspace{10pt}

\section{Introduction}\label{sec:introduction}

Wireless power transfer (WPT) is an emerging technology set to supply wireless devices in the future generation of self-sustaining and battery-free Internet of Things (IoT) and wireless sensor networks (WSN) \cite{kim2021wireless}. 
RF transmission is widely used for wireless data transfer and wireless communications. 
Additionally, RF waves can also carry energy together with information. 
Due to the recent improvement in power consumption of electronics, leading to miniaturization and reduction of power consumption of IoT devices, WPT is increasingly emerging as a potential and feasible solution as power supply technology for them \cite{meile2019wireless, Hemour2014, Smith2013}.

Wireless power transfer has several advantages over environmental energy harvesting for the supply of IoT devices \cite{kim2021wireless}. 
The main advantage is the controllable nature of wireless power transfer versus the unpredictable nature of environmental energy harvesting \cite{meile2020radio}. 
In fact, instead of harvesting energy for environmental sources, WPT technology is designed to exploit a stable and controllable wireless power supply by deploying a dedicated power transmitter that periodically or continuously sends energy
\cite{Clerckx2019}. 
Moreover, RF WPT allows long-distance energy delivery to a multitude of IoT nodes and enables a small form factor at the receiver with compact receiving antennas, especially when compared to other technologies such as inductive coupling or magnetic resonance.
For these reasons, the interest for RF WPT has increased in research and industry in recent years \cite{meile2020radio,  PowercastCorp2021}.
The vision for this emerging technology is to permanently power low-power devices, eliminating the need for batteries or increasing their lifespans and reducing maintenance without loss of performance \cite{Martins2021}. 
This can be especially attractive in situations where a periodical change of batteries is difficult or not desired. For these reasons, powering the sensors through wireless power sources provides an effective alternative to other energy harvesting solutions such as solar, thermal, or vibration \cite{clerckx2022future}.

WPT is mainly divided into near- and far-field WPT depending on the spatial range of the transfer. 
Near-field inductive WPT is used to wirelessly power many commercially available near-field communication (NFC) devices such as credit cards and smartphones and is widely adopted \cite{lathiya2021near}.
However, its range is limited to few a centimeters, due to the mainly inductive transfer of energy and data in the near-field. 
On  the contrary, far-field WPT is based on sending and receiving electromagnetic (EM) waves using dipoles, which allows for power transmission at longer ranges of over 1 meter \cite{ayestaran2019wireless}. 
This makes far-field WPT more flexible and particularly suitable for power multi-casting, allowing the transmitter/receiver to move around, even in Non-Line-of-Sight (NLoS) environments \cite{huang2019wireless}. 
Support of longer ranges is crucial to supply energy to IoT devices in a wide range of applications \cite{ayestaran2019wireless }. 
As a result, there has been a fair amount of research dedicated to the design of dedicated WPT transmitters for these sensor systems \cite{ayestaran2019wireless,meile2019wireless,meile2020radio}. 
Despite this, far-field RF-WPT over longer ranges is still very challenging. One of the major challenges is the efficient transformation of the input EM power levels smaller than \SI{-10}{dBm}, to the voltage and current demands of the load \cite{Park2020}. This is exacerbated by the need to use a miniaturized antenna and compact designs for small and unobtrusive  IoT devices \cite{Paolini2021}.

RF WPT is also increasing its popularity due to the availability of novel commercial products. For instance, Powercast allows power transfer up to \SI{200}{\milli\watt} using a single narrow-band sub-GHz channel over several meters. 
Having power available at this range, combined with the possibility to supply several receivers from a single transmitter, makes RF WPT a promising power source for low-power IoT devices \cite{meile2019wireless}. 
However, it must be taken into account that the effective power available at the IoT nodes is strongly affected by the distance, antenna size, RF matching, and rectification efficiency, and will therefore be orders of magnitude below the transmitted power of \SI{200}{\milli\watt}. Therefore, to efficiently extract power from EM waves, a carefully designed RF EH system is of utmost importance.

This paper presents the electronic blocks of an energy-efficient RF WPT sub-system for miniaturized IoT and focuses on critical design considerations for each part. The design and evaluation of an optimized RF power rectifier and matching circuit for \SI{915}{\mega\hertz} for input power levels ranging from \SI{-12}{dBm} to \SI{9}{dBm} is presented. 
Experimental results demonstrate that the optimized design increases the conversion efficiency by up to \SI{20}{\percent} compared to off-the-shelf circuitry. Finally, the proposed sub-system, but also other commercially available RF rectifiers and power management systems are exploited to maximize the PCE. 
Experimental results show that is possible to achieve an overall PCE of over \SI{30}{\percent} for an input power of \SI{-10}{dBm} which is above the state of the art for such systems.
{\color{blue}{
}}

\section{Architecture}\label{sec:systemarchitecture} 

The goal of a WPT sub-system is to extract power from electromagnetic waves received by the antenna as efficiently as possible to achieve power transfer over longer distances. Eventually, the captured electromagnetic energy is converted into a usable direct current (DC) voltage.
To achieve this, each of the three main blocks of the EH system (ref. Fig. \ref{fig:custom_rf} (a)), 
impedance matching network, rectifier, and power management with energy storage requires a carefully considered design.

\subsection{Impedance Matching Network}\label{sub:impedance_matching}
To achieve maximal power transfer between the source (antenna) and the load (rectifier), a carefully designed impedance matching network (IMN) is essential. By transforming the low impedance of the antenna to the system impedance, power reflections can be mitigated. Choosing discrete components with a high-quality factor minimizes power dissipation and further improves the IMN's conversion efficiency \cite{Clerckx2019}. 
\subsection{Rectifier}\label{sub:rectifier}
When designing a high efficient RF-to-DC rectifier, selecting appropriate diodes is one of the most important factors since they are the main source of loss and their characteristic determines the overall performance of the rectifier.
Zero-bias Schottky diodes provide very low threshold voltage and junction capacitance in comparison to common diodes, making them well-suited for RF rectifiers and allowing efficient operation \cite{Clerckx2019}.

\subsection{Power Management with Energy Storage}\label{sub:power_management}
\begin{figure}
    \centering
    \begin{overpic}[width=0.95\columnwidth]{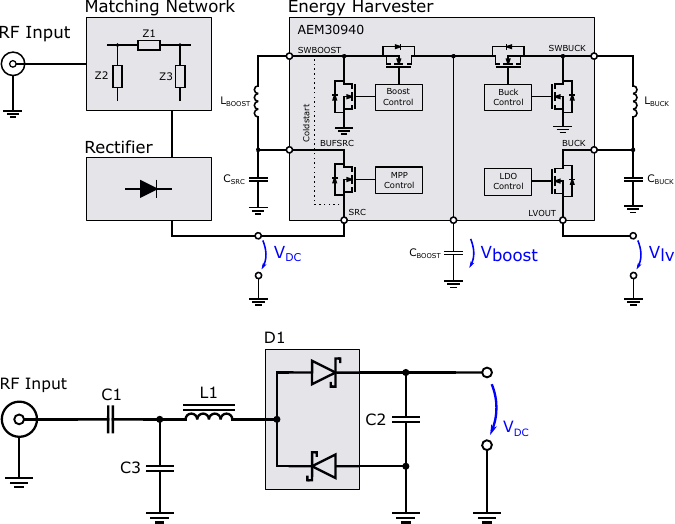}
        \put(0,78){(a)}
        \put(0,26){(b)}
    \end{overpic}
	\caption{(a) High-level circuit block diagram, (b) schematic of the proposed and implemented RF rectifier and matching circuit. Selected components values are listed in Table \ref{table:selected_components}.}
	\label{fig:custom_rf}
\end{figure}

After rectification, the harvested RF power is available as DC power and can be already used to drive a load. However, if the load's power needs cannot be guaranteed at all times, which is especially true for miniaturized IoT devices, power management with energy storage is needed \cite{Martins2021}. By applying strict power-saving techniques to the IoT devices, the system can still achieve self-sustainability \cite{Magno2019}.
\newline
The proposed WPT sub-system, illustrated in Fig. \ref{fig:custom_rf} (b), is optimized for a single narrow-band RF power transmitter with a center frequency of \SI{915}{\mega\hertz} and is realized using off-the-shelf components. 
High-Q capacitors and inductors of the series \textit{GJM1555} and \textit{LQW18AN} keep the conversion losses at a minimum. 
C\textsubscript{1} is a DC block to allow the rectifying circuits' DC bias to settle at mid-rail voltage. 
C\textsubscript{2} and L\textsubscript{1}, together with the diode capacitances, build a single band matching network in $\pi$-configuration. The rectifier structure itself is based on a voltage doubler using \textit{BAT15-04W} zero-bias RF Schottky diodes for efficient RF-to-DC conversion.

High-quality RF matching requires a well-defined printed circuit board (PCB) stackup. Thus, the proposed circuit has been realized on a 2-layer PCB using RO4003c as the core dielectric with a thickness of \SI{0.813}{\milli\meter}. RO4003C provides a low and controlled dielectric constant of 3.38 over a wide frequency range and a dissipation factor of 0.0027 at \SI{10}{\giga\hertz}, making it well suited for power-sensitive high-frequency designs.

For power management and energy storage, the \textsc{e-peas} AEM30940 energy management circuit has been selected [ref. Fig.\ref{fig:custom_rf} (a)]. It can start operating with empty storage elements from an input voltage as low as \SI{380}{\milli\volt} and an input power of \SI{3}{\micro\watt}. Once active, power from input voltages as low as \SI{50}{\milli\volt} can be converted and temporarily stored in a capacitor before generating a constant output voltage level with the internal buck converter and low drop out (LDO) regulator.

\begin{table}[!ht]
    \caption{Selected components of the proposed circuit}
    \centering
    \renewcommand{\arraystretch}{1.35}
    \begin{tabular}{@{}l*{2}{c}c@{}}
        \toprule
        Designator & \hspace{30px} Series \hspace{30px} & Value \\
        \midrule
        C1  & GJM1555   & \SI{33}{\pico\farad}  \\
        C2  & GQM1875   & \SI{2.2}{\pico\farad} \\
        C3  & GJM1555   & \SI{2.4}{\pico\farad} \\
        L1  & LQW18AN   & \SI{50}{\nano\henry}  \\
        D1  & -         &  BAT15-04W            \\
        \bottomrule
    \end{tabular}
    \label{table:selected_components}
\end{table}
\begin{figure*}[!t]
    \centering
    \begin{overpic}[width=1\linewidth]{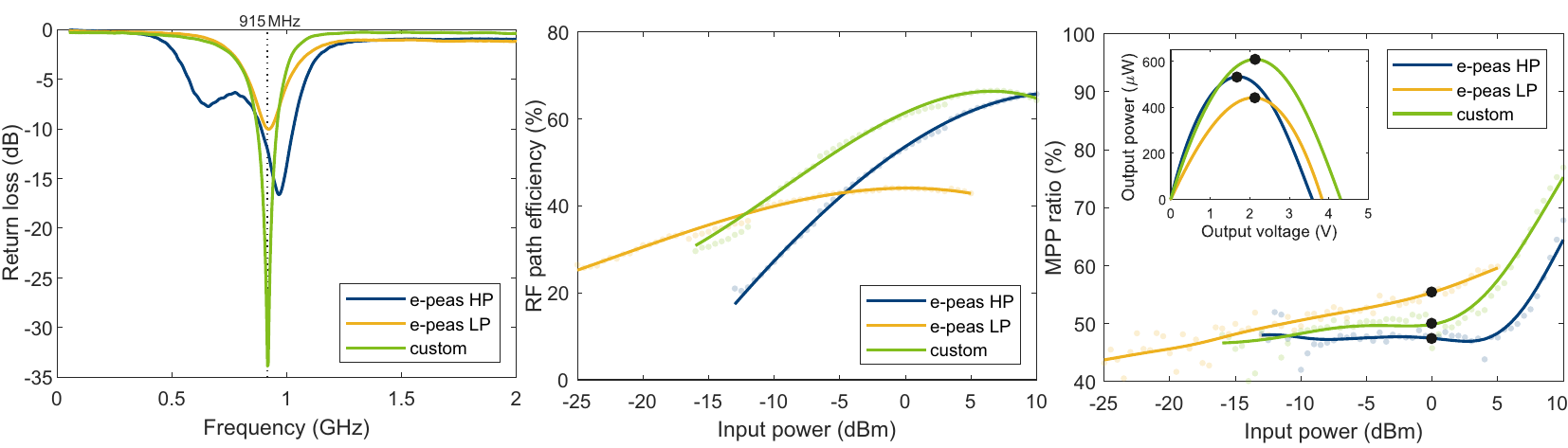}
        \put(0,28){(a)}
        \put(33,28){(b)}
        \put(66,28){(c)}
    \end{overpic}
	\caption{Characterization of the matching and RF rectification circuit. (a) S11 return loss of the three analyzed circuits for \SI{0}{dBm} input power. (b) Rectification efficiency as a function of the input power with optimally matched loads. (c) Optimal maximal power point as a ratio of the open circuit voltage. The inset visualizes the output power for varying load conditions at \SI{0}{dBm} input power.}
	\label{fig:rectifier}
\end{figure*}

\section{Experimental Results}\label{sec:results}
To evaluate the performance of the proposed sub-system it has been tuned for \SI{915}{\mega\hertz} which is a commonly used carrier frequency for WPT \cite{meile2019wireless,PowercastCorp2021}. The individual functional blocks have been evaluated in terms of functionality and efficiency. Aside from the proposed narrow-band rectification circuit of section \ref{sec:systemarchitecture}, measurements are conducted on the \textsc{e-peas} AEM30940 evaluation board to facilitate reproducibility.

The evaluation of the matching and rectifier characteristics, including an optimal maximal power point configuration analysis, is given in subsection \ref{subsec:rectifier}. Subsection \ref{subsec:fullsystem} evaluates the end-to-end harvesting efficiency of the e-peas harvesting integrated circuits (IC) when combined with different RF frontends. Finally, subsection \ref{subsec:coldstart} analyzes the cold-start sequence of the harvesting IC for low input power.

\subsection{Matching Circuit and Rectifier}\label{subsec:rectifier}
Ideally, RF harvesting circuits comprise a broadband antenna and non-frequency selective matching circuits, allowing them to harvest from a wide range of carrier signals and input powers. If the application, the carrier frequency, and the typical input power is known, the RF path can be matched precisely to the input impedance of the rectifying circuit. 

\cref{fig:rectifier} visualizes the characterization of the matching and rectification circuitry. For the evaluation, the two selectable input frontends on the AEM30940 evaluation board, \emph{e-peas HP} and \emph{e-peas LP}, respectively optimized for power levels above \SI{-10}{dBm}  and below \SI{-10}{dBm}, are compared against the narrow band implantation of section \ref{sec:systemarchitecture} "custom". The circuit matching is analyzed with an HP7853 network analyzer at a constant source power of \SI{0}{dBm} and a matched resistive load on the rectifier output replacing the subsequent circuitry. \cref{fig:rectifier} (a) shows the S11 parameter of the three circuits for frequencies of up to \SI{2}{\giga\hertz}. The 
high-quality factors of the selected components allow a narrow matching to the \SI{915}{\mega\hertz} target frequency for the \emph{custom} rectifier. In contrast, the \emph{e-peas HP} circuit allows additional harvesting from the \SI{868}{\mega\hertz} band.

To precisely analyze the rectification efficiency of the previously shown matching and rectifying circuits, an \textsc{R\&S} SML03 signal generator configured to \SI{915}{MHz} is connected to the circuit input. The peak output power and the corresponding maximal power point (MPP) configuration are measured with a source measurement unit (SMU) \textsc{Keysight} B2901A emulating an adaptive load. \cref{fig:rectifier} (b) shows the resulting RF path efficiency as a function of the input power for optimally matched loads. The results suggest that for input signals higher than \SI{-5}{dBm}, the \emph{e-peas HP} outperforms the \emph{e-peas LP} configuration. In contrast, the custom implementation shows superior performance for input power levels between \SI{-12}{dBm} and \SI{9}{dBm} with up to \SI{20}{\percent} efficiency gain as compared to the commercial circuit's efficiency. The corresponding MPP configuration is shown in \cref{fig:rectifier} (c). The \emph{e-peas LP} configuration shows a linear shift with \SI{0.4}{\percent/dBm} over the analyzed power range. The circuits designed for higher input powers are approximately stable until  \SI{4}{dBm}, with a significantly rising  MPP ratio above. The inset visualizes the output power for a varying load at \SI{0}{dBm}.



\subsection{Wireless Power Conversion Efficiency}\label{subsec:fullsystem}
Correctly controlling the output voltage of the RF path is essential to efficiently supply subsequent circuitry. The circuit behavior, when combined with an \textsc{epeas} AEM30940 harvesting IC, is shown in \cref{fig:fullSystem} (b). For the measurement, the previously tested RF frontends have been  attached to the IC's harvesting input. To allow a measurement of the end-to-end harvesting efficiency, shown in \cref{fig:fullSystem} (a) an SMU configured to \SI{3.5}{\volt} has been connected in parallel to an intermittent energy storage capacitor $C_{boost}$. This allows monitoring of the harvested current at a stable and repeatable storage configuration. The embedded maximal power point tracking (MPPT) has been configured to \SI{50}{\percent} of the open circuit voltage. 
The \emph{e-peas LP} frontend allows energy-positive operation of the harvesting circuitry down to \SI{-15}{dBm}. Combined with the high power optimized frontends, the circuit reaches \SI{-16}{dBm} and \SI{-15}{dBm} for \emph{custom} and \emph{e-peas HP}, respectively. The custom circuit combined with the AEM30940 IC shows increased harvesting efficiency for input power levels between \SI{-11}{dBm} and \SI{5}{dBm} with a peak efficiency of \SI{57}{\percent} at \SI{3}{dBm}.

\subsection{Cold Start}\label{subsec:coldstart}
\begin{figure*}[!t]
    \centering
    \begin{overpic}[width=1\linewidth]{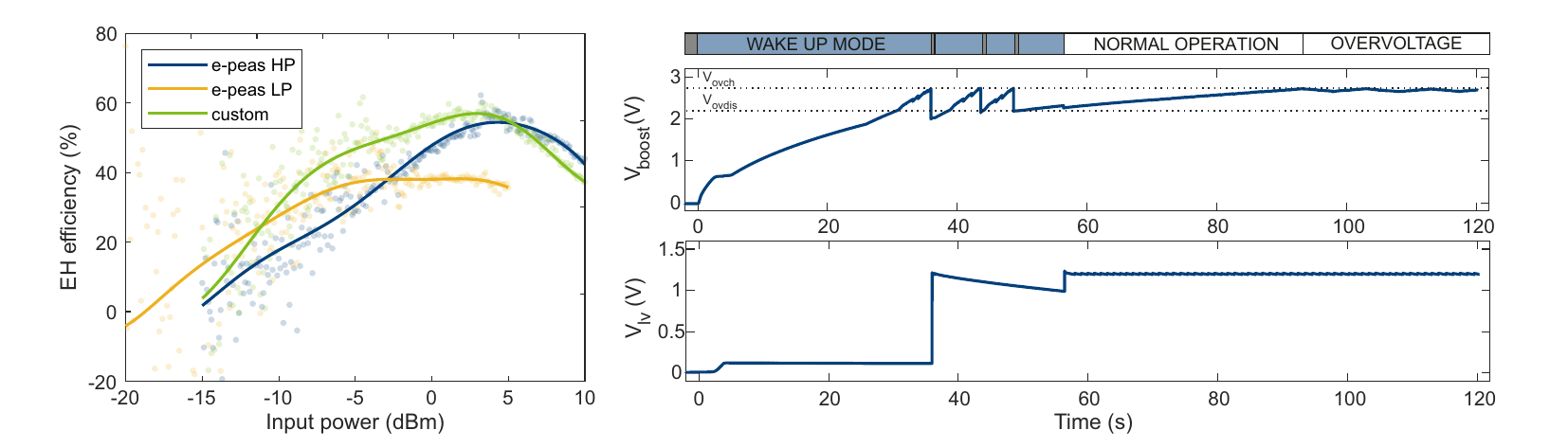}
        \put(3,28){(a)}
        \put(39,28){(b)}
    \end{overpic}
	\caption{AEM30940 harvesting performance: (a) End-to-end harvesting efficiency for a constant boost converter output voltage of \SI{3.5}{\volt}. (b) Cold-start sequence of the AEM30940 IC hosting the custom RF frontend at an input power of \SI{-15}{dBm}.}
	\label{fig:fullSystem}
\end{figure*}

A critical feature of intermittent operation devices is the capability to power up despite totally discharged energy storage. \cref{fig:fullSystem} (b) shows this cold-start behavior for the AEM30940 IC connected to the RF frontend presented in section \ref{sec:systemarchitecture}. The IC's overcharge protection has been configured to $V_{ovch}=\SI{2.7}{\volt}$ and the under-voltage lockout level to $V_{ovdis}=\SI{2.2}{\volt}$. The regulator output voltage is $V_{lv}=\SI{1.2}{\volt}$. 

At $t=\SI{0}{\second}$, the harvesting circuit is supplied with a \SI{915}{MHz} RF input signal with a power of \SI{-15}{dBm}. After \SI{35}{\second}, the initial wake-up phase completes as the intermittent storage capacitor $C_{boost}$ charges to the programmed maximum. The activation of the output regulator $V_{lv}$ and the corresponding inrush currents combined with the limited input power results in consecutive under-voltage lockouts until normal operation is reached after \SI{56}{\second}. Finally, after \SI{93}{\second}, $C_{boost}$ is fully charged, and the circuit operates in overcharge protection mode. 

\section{Conclusion}\label{sec:conclusion}
The paper proposed and evaluated a WPT sub-system, composed of an impedance matching network, rectifier, and power management with energy storage, optimized to a single narrow-band channel at \SI{915}{\mega\hertz}. The experimental evaluation showed an overall PCE of over \SI{30}{\percent} at an input power of \SI{-10}{dBm} and a peak efficiency of \SI{57}{\percent} at \SI{3}{dBm}. In comparison to commercial solutions, the proposed RF frontend showed an increased PCE for input power levels between \SI{-11}{dBm} and \SI{5}{dBm}.
The commercial IC AEM30940 from \textsc{epeas} allows to temporarily provide more power than is being harvested, acting as a buffer and keeping the voltage supply constant, which is especially important for miniaturized IoT devices. Using the proposed RF frontend, the AEM30940 is able to exit cold start after \SI{56}{\second} from an input power as low as \SI{-15}{dBm} and switches to normal operation.


\bibliography{./main}

\end{document}